\documentclass[12pt]{iopart}
\usepackage{iopams}
\usepackage{epsfig}
\usepackage{textcomp}
\usepackage{citesort}
\begin{document}

\title[Does strong heterogeneity promote cooperation by group interactions?]{Does strong heterogeneity promote cooperation by group interactions?}

\author{Matja{\v z} Perc}
\address{Department of Physics, Faculty of Natural Sciences and Mathematics, University of \\ Maribor, Koro{\v s}ka cesta 160, SI-2000 Maribor, Slovenia}
\ead{matjaz.perc@uni-mb.si}

\begin{abstract}
Previous research has highlighted the importance of strong heterogeneity for the successful evolution of cooperation in games governed by pairwise interactions. Here we determine to what extent this is true for games governed by group interactions. We therefore study the evolution of cooperation in the public goods game on the square lattice, the triangular lattice and the random regular graph, whereby the payoffs are distributed either uniformly or exponentially amongst the players by assigning to them individual scaling factors that determine the share of the public good they will receive. We find that uniformly distributed public goods are more successful in maintaining high levels of cooperation than exponentially distributed public goods. This is not in agreement with previous results on games governed by pairwise interactions, indicating that group interactions may be less susceptible to the promotion of cooperation by means of strong heterogeneity as originally assumed, and that the role of strongly heterogeneous states should be reexamined for other types of games.
\end{abstract}

\maketitle

\section{Introduction}

One possible classification of evolutionary games \cite{sigmund_93,hofbauer_98,nowak_06,sigmund_10} is to those that are governed by pairwise interactions and to those that are governed by group interactions. The prisoner's dilemma and the snowdrift game \cite{axelrod_84,poundstone_92}, as well as the stag-hunt game \cite{skyrms_04}, are classical examples of evolutionary games that are governed by pairwise interactions. The public goods game, on the other hand, is a typical example of an evolutionary game that is governed by group interactions. Similarly as by the transition from two-body to many-body interactions in many branches of physics, in evolutionary games too the transition from pairwise to group interactions leads to a substantial increase in complexity. For example, reciprocity \cite{trivers_qrb71,nowak_n98}, \textit{i.e.} the act of returning favor for a favor, is straightforward in games governed by pairwise interactions. As there are only two players involved at each instance of the game, it is relatively easy to decide what to do based on what the opponent has done in the past \cite{baek_pre08}. In games governed by group interactions, however, it is much more difficult to keep track of actions of all the other players, and hence it is difficult to reciprocate. The same argument is valid for punishment \cite{sigmund_tee07}, where unlike returning positive actions, the goal is to identify those that inflict harm or act antisocial and sanction them accordingly.

Regardless of the distinction between pairwise and group interactions, the central theme of evolutionary game theory is the successful evolution of cooperation. According to Darwin, natural selection favors the fittest and the most successful individuals, which implies an innate selfishness that greatly challenges the concept of cooperation \cite{axelrod_84}. To cooperate namely means to sacrifice some fraction of personal benefits for the sake of social welfare. The opposite strategy is defection, implying that players who defect will always try to maximize their fitness regardless of the consequences this might have for the society. Since the focus here is on games governed by group interactions, we may use the definition of the public goods game to illustrate the key difference between cooperation and defection as follows. In a group consisting of $G$ players, those that cooperate will contribute $1$ to the public good, while those that defect will contribute nothing. Note that on a larger scale, the contribution of cooperators can be considered as a contribution to society, which will typically contain a large number of such groups. All the contributions will then be multiplied by a factor $r$, which is typically larger than $1$ to take into account synergetic effects of collaborative efforts, and subsequently the resulting amount will be shared amongst all the group members irrespective of their initial contribution. From this definition it follows that defectors acquire an evolutionary advantage over cooperators by withholding the initial contribution to the public good, and from the viewpoint of each individual defection is thus clearly the rational strategy to choose. From the viewpoint of the group and the society as a whole, however, cooperation is the optimal strategy as then the multiplication factor $r$ will have the biggest impact and accordingly the welfare of the society will be maximized. The fact that defection maximizes individual payoffs while cooperation maximizes social prosperity, and thus that what is best for an individual is opposite to what is the best for society, is traditionally refereed to as a social dilemma. Failure to maintain cooperation in such a case leads to the so-called ``tragedy of the commons'' \cite{hardin_g_s68}, where nobody contributes to the public good (everybody defects) and the society is therefore destined to go bankrupt. Rather surprisingly, and in fact contradictive to the Darwinian concept of acting so as to maximize personal fitness, economic experiments on public goods games indicate that humans cooperate much more in such situations than expected \cite{fehr_ars07}, and in so doing these experiments call for the identification of mechanisms that can explain the successful evolution of cooperation.

Introducing additional strategies besides cooperation and defection has proven to be very effective. For example, if players are given the chance to abstain from the public goods game by not contributing but also by not taking part in the distribution of accumulated payoffs, cooperation may be promoted by means of the spontaneous emergence of cyclic dominance between the three competing strategies \cite{hauert_s02,szabo_prl02,semmann_n03}. The introduction of peer-punishers, \textit{i.e.} those that are willing to bare additional costs in order to sanction defectors, has also proven to be effective for the promotion of cooperation \cite{hauert_s07,helbing_njp10}, as was the introduction of pool-punishers that contribute to the establishment of sanctioning institutions \cite{sigmund_n10,szolnoki_pre11}. Rewarding cooperative players instead of punishing defectors has also been considered as an additional strategy that may promote cooperation \cite{rand_s09,hauert_jtb10,szolnoki_epl10}, and the emerging ``stick versus carrot'' dilemma (whether to sanction defectors or reward cooperators) has recently received ample attention \cite{hilbe_prsb10}. Interestingly, random explorations of these additional strategies, termed conveniently as ``exploration dynamics" \cite{traulsen_pnas09}, may also substantially elevate the level of cooperation in a society.

Driven by the application of methods from statistical physics (see \cite{szabo_pr07} for a comprehensive review), as well as by the fascinating complexity arising from the evolutionary competition between the strategies, physicists have also made important contributions to the understanding of the successful evolution of cooperation. Foremost, research published in recent years has made it clear that heterogeneities amongst players play a crucial role by the evolution of cooperation. Scale-free networks, for example, have been recognized as very potent promoters of cooperative behavior \cite{santos_prl05,gomez-gardenes_prl07,poncela_njp07,santos_n08}. In fact, evolutionary games on complex and coevolving networks \cite{abramson_pre01,ebel_pre02,zimmermann_pre04,lozano_ploso08,vukov_pre08,poncela_njp09,perc_bs10} in general tend to promote cooperation past the boundaries imposed by regular lattices \cite{nowak_n92b,szabo_pre98}. Similarly, heterogeneities in strategy adoption probabilities can also enhance cooperation \cite{kim_bj_pre02,wu_zx_pre06}, especially if the strategy adoption is favored from the more influential players \cite{szolnoki_epl07,perc_pre08b}. Heterogeneities can also be introduced directly to payoffs in terms of noise \cite{perc_njp06a,vukov_pre06,tanimoto_pre07b} or quenched diversity \cite{perc_pre08}, whereby cooperators are promoted as well provided the uncertainties are adequately adjusted and distributed.

While the majority of previous works aimed at disentangling the impact of heterogeneity on the evolution of cooperation focused on games governed by pairwise interactions, recent results indicate \cite{wang_j_pre09,wu_t_epl09,rong_epl09,shi_dm_epl10,wang_j_pre10b,gomez-gardenes_c11} that cooperation within the public goods game is also susceptible to the same mechanism of promotion. However, if payoffs are evaluated from public goods games in multiple groups, the indirect linkage of players due to their membership in the same groups may result in qualitatively different behavior \cite{szolnoki_pre09c} as reported previously for games governed by pairwise interactions \cite{szabo_pre05}. This motivates us to examine to what extent strongly heterogeneous states do in fact promote the evolution of cooperation in games governed by group interactions. For this purpose, we study the evolution of cooperation in the public goods game on three types of regular graphs, namely on the square lattice, the triangular lattice, and the random regular graph (RRG). These are characteristic for interaction graphs with (square and triangular lattice) and without (RRG) spatial structure, as well as for interaction graphs with zero (square lattice and RRG) and a high (triangular lattice) clustering coefficient, thus covering a broad plethora of properties that are known to vitally affect the evolution of cooperation \cite{szabo_pr07}. Heterogeneity is then introduced by means of either a uniform or an exponential distribution of payoffs amongst the players, thereby violating the traditional assumption of equally distributed public goods amongst all group members. According to the statistical properties of the two considered distributions, exponentially distributed public goods give rise to strongly heterogeneous states, while uniformly distributed public goods correspond to moderate heterogeneities affecting the evolution of cooperation. Unexpectedly, we find that moderate heterogeneities promote cooperation better than strongly heterogeneous states, which is not in agreement with results obtained previously for evolutionary games governed by pairwise interactions \cite{perc_pre08}.

The remainder of this paper is organized as follows. In the next section we describe the employed evolutionary public goods game and the interaction graphs, while in section 3 we present the results. Lastly, we summarize our findings and compare them with those reported earlier for evolutionary games governed by pairwise interactions.

\section{Setup}

\begin{figure}
\begin{center} \includegraphics[width = 8cm]{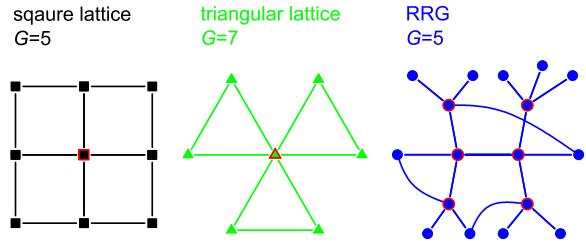}
\caption{\label{graphs} Schematic presentation of interaction graphs considered in this paper. While the square lattice (left) and the random regular graph (right) both contain groups with $G=5$ players each, the triangular lattice (middle) has $G=7$. Only vertices encircled red have all their neighbors depicted.}
\end{center}
\end{figure}

Assuming structured interactions defined by either the square lattice, the triangular lattice, or the random regular graph, as schematically depicted in figure~\ref{graphs}, $L^2$ players are arranged into overlapping groups of size $G$ such that every player is surrounded by its $G-1$ neighbors. Accordingly, each individual belongs to $g=G$ different groups. Initially each player on site $x$ is designated either as a cooperator ($s_x = C$) or defector ($s_x = D$) with equal probability. Cooperators contribute a fixed amount (here considered being equal to $1$ without loss of generality) to the public good while defectors contribute nothing. The sum of all contributions in each group is multiplied by the factor $r$ and the resulting public goods are distributed amongst all the group members. If $s_x = C$ the payoff of player $x$ from every group $g$ is $P_{C}^g=(1+h_x) r N_{C}^g/G - 1$ and if $s_x = D$ the payoff is $P_{D}^g=(1+h_x) r N_{C}^g/G$, where $N_{C}^g$ is the number of cooperators in group $g$ while $h_x$ is the scaling factor by means of which the heterogeneity in the distribution of public goods is introduced. The scaling factors are drawn randomly from either the uniform distribution $h = \eta (-2 \chi + 1)$ or the exponential distribution $h = \eta (-\log \chi - 1)$. Here $\chi$ are uniformly distributed random numbers from the unit interval, and $\int_0^1 h(\chi) d\chi = 0$ in all cases, so that the average $L^{-2}\sum h_x$ over all the players is zero. Moreover, $\eta$ scales the magnitude of heterogeneity and will, together with $r$, be considered as the key parameter affecting the evolution of cooperation. Naturally, $\eta=0$ returns the traditional version of the game, while large $\eta$ lead to segregation of players, which may be additionally amplified by considering exponential rather than uniformly distributed $h_x$. It should be noted that heterogeneity is here not quantified explicitly (\textit{e.g.} by means of standard deviation of $h_x$), but rather it refers to the diversity of players in the sense of to what extend they can differ from one another. Since the exponential distribution segregates the players much more than the uniform distribution, the former is referred to as being substantially more heterogeneous (see also \cite{perc_pre08} for explicit shapes of the two distributions and alternative ways of interpretation).

Related to this setup, it is important to note that quenched heterogeneities introduced via $h_x$ may evoke the existence of the Griffiths phase \cite{griffiths_prl69}, which has recently attracted considerable attention \cite{munoz_prl10,vazquez_prl11}, also in studies concerning the evolution of cooperation \cite{droz_epjb09}. The essence of the problem of quenched heterogeneities for the extinction processes has been well described in \cite{noest_prl86,noest_prb88}, where it was shown that such systems are frequently characterized by patches of different sizes, providing better conditions for one of the strategies (or species) to survive. Due to the localization, the subordinate strategy can die out very slowly on the separated (or weakly interacting) patches, with an average lifetime increasing with the patch size. Noest \cite{noest_prl86,noest_prb88} demonstrated that for suitable conditions (determined by the distribution of patch sizes) the extinction of the subordinate strategy follows a power law, whereby the exponent depends on the parameters. The latter fact can cause serious technical difficulties in the classification of the final stationary state, especially related to the $C+D \to C$ transition in game theoretical models, as demonstrated for example in figure~3 of \cite{droz_epjb09}, where it can be inferred that even very long simulation times might not be enough to reach the final stationary state, although the trend (power law behavior) clearly indicates the disappearance of the subordinate strategy in the limit when the time goes to infinity. We note that often the introduction of an additional time-dependence in the background can make the analysis more convenient. For example, if the quenched heterogeneities are varied on an extremely slow time scale (much slower than is characteristic for the main evolutionary process), the final conclusions remain the same, yet the occasional variations can accelerate the extinction significantly. This approach seems viable for alleviating difficulties that are frequently associated with models entailing quenched heterogeneities.

Monte Carlo simulations are carried out comprising the following elementary steps. First, a randomly selected player $x$ plays the public goods game with its $G$ partners as a member of all the $g=1, \ldots, G$ groups. The payoff it thereby acquires is thus $P_{s_x} = \sum_g P_{s_x}^g$. Next, player $x$ chooses one of its nearest neighbors at random, and the chosen co-player $y$ also acquires its payoff $P_{s_y}$ in the same way as player $x$. Finally, player $x$ enforces its strategy $s_x$ onto player $y$ with the probability $w(s_x \to s_y)=1/\{1+\exp[(P_{s_y}-P_{s_x})/GK]\}$, where $K=0.1$ quantifies the uncertainty by strategy adoptions and $G$ normalizes the effect for different interaction graphs (see \cite{szolnoki_pre09c} for details concerning the selection of the $K$ value and the applied normalization). These three elementary steps [1 -- random selection of player $x$ and the determination of $P_{s_x}$, 2 -- random selection of neighbor $y$ and the determination of $P_{s_y}$, 3 -- attempted strategy transfer with probability $w(s_x \to s_y)$] are repeated consecutively, whereby each full Monte Carlo step (MCS) gives a chance for every player to enforce its strategy onto one of the neighbors once on average and is comprised of $L^2$ three elementary steps described above. The average fractions of cooperators ($\rho_{C}$) and defectors ($\rho_{D}$) in the population were determined in the stationary state after sufficiently long relaxation times. Depending on the actual conditions (proximity to extinction points and the typical size of emerging spatial patterns) the linear system size was varied from $L=200$ to $800$ and the relaxation time was varied from $10^4$ to $10^6$ MCS to ensure proper accuracy.

\section{Results}

\begin{figure}
\begin{center} \includegraphics[width = 12cm]{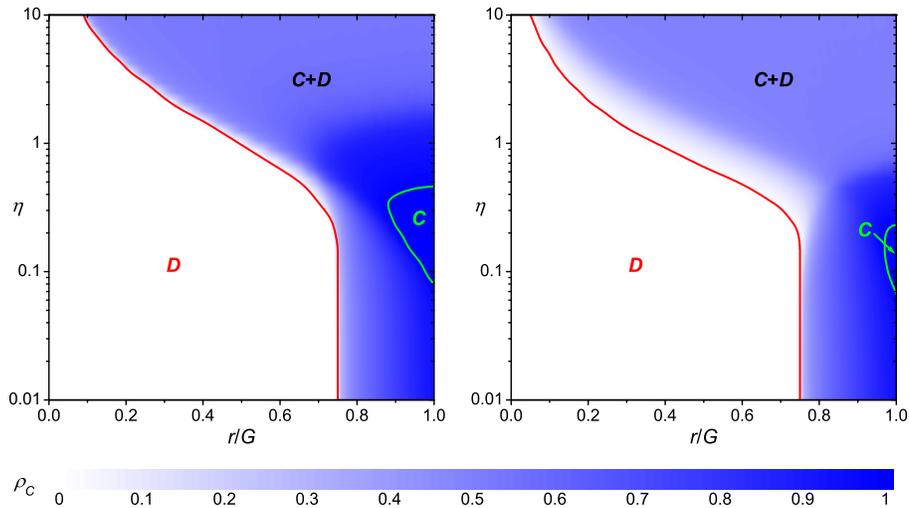}
\caption{\label{sqr} Full $r-\eta$ phase diagrams for the square lattice with uniformly (left) and exponentially (right) distributed public goods. Red and green lines delineate the $D \to C+D$ and $C+D \to C$ phase transitions, respectively. The blue shading corresponds to the stationary density of cooperators $\rho_C$ at each particular combination of $r$ and $\eta$, as depicted by the color bar at the bottom. It can be observed that uniformly distributed public goods (left) promote cooperation better than the more heterogeneous exponentially distributed public goods (right). See also main text for details.}
\end{center}
\end{figure}

We start by presenting results obtained on the square lattice with $G=5$ (depicted schematically in figure~\ref{graphs} left) in figure~\ref{sqr}. The left panel features the full $r-\eta$ phase diagram for uniformly distributed public goods. Focusing on the $D \to C+D$ transition line (depicted red), marking the survivability threshold of cooperators, we find that increasing values of $\eta$ continuously decrease the minimally required multiplication factor $r$. From $r=3.74$ at $\eta=0.01$ and $\eta=0.1$ (note that such weekly expressed heterogeneity actually returns the traditional version of the game where the public goods are distritbuted equally \cite{szolnoki_pre09c}), the minimally required multiplication factor decreases to $r=2.45$ at $\eta=1.0$, and further to $r=0.45$ at $\eta=10$. Importantly, note that such contradictory critical values of $r$ that are smaller than $1$ are possible only when individual $h_x$ can exceed $G$, \textit{i.e.} when some individuals can effectively collect shares of the public good also from players that are not within the same group. For the uniformly distributed public goods it is straightforward to determine that this happens when $\eta=G$, which is also approximately where the critical multiplication factor $r$ dips below $1$. For such high $\eta$, however, the public goods game is essentially overridden and the stationary state is always $\rho_{C} \approx \rho_{D}$, irrespective of $r$ (within the mixed $C+D$ phase). The pure $C$ phase, \textit{i.e.} $\rho_{C}=1$, on the other hand, emerges only for intermediate values of $\eta$, as depicted by the green line marking the $C+D \to C$ transition.

By comparing the results depicted in the left panel of figure~\ref{sqr} with those depicted in the right panel, we can observe that exponentially distributed public goods (right) on the square lattice fail to promote cooperation to the same extent as uniformly distributed public goods (left). While the position of the $D \to C+D$ transition line remains largely unaffected, the mixed $C+D$ phase is characterized by substantially lower values of $\rho_C$ (brighter shades of blue prevail). Moreover, the pure $C$ region is substantially smaller, and the override of the public goods game, where $\rho_{C} \approx \rho_{D}$ irrespective of $r$ (within the mixed $C+D$ phase) sets in already for $\eta \gtrsim 0.7$ (see left panel of figure~\ref{sqr}). These observations indicate that strongly heterogeneous states (constituted by exponentially distributed public goods) may be less effective in warranting high levels of cooperation at low multiplication factors than moderately heterogeneous states (constituted by uniformly distributed public goods). Irrespective of the type of heterogeneity, however, the survivability of cooperators can be greatly enhanced, \textit{i.e.} the minimally required $r$ for $\rho_{C}$ to be larger than zero greatly reduced.

\begin{figure}
\begin{center} \includegraphics[width = 12cm]{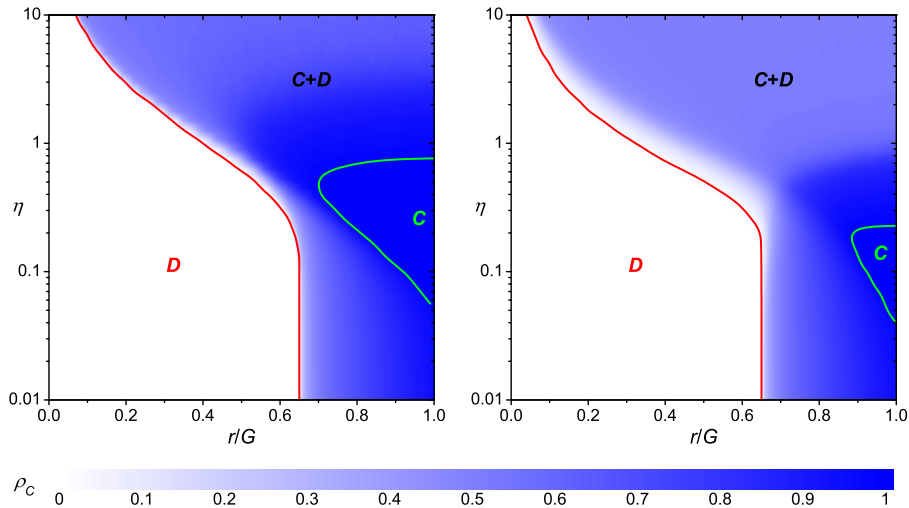}
\caption{\label{tri} Full $r-\eta$ phase diagrams for the triangular lattice with uniformly (left) and exponentially (right) distributed public goods. Red and green lines delineate the $D \to C+D$ and $C+D \to C$ phase transitions, respectively. The blue shading corresponds to the stationary density of cooperators $\rho_C$ at each particular combination of $r$ and $\eta$, as depicted by the color bar at the bottom. Even if interaction graphs with a high clustering coefficient are considered (such as the triangular lattice), uniformly distributed public goods (left) still outperform exponentially distributed public goods (right) in terms of the promotion of cooperation. See also main text for details.}
\end{center}
\end{figure}

Figure~\ref{tri} features the results of the same analysis as depicted in figure~\ref{sqr}, only that instead of the square lattice the triangular lattice was used as the interaction graph. The use of the triangular lattice is motivated by its high clustering coefficient (note that the square lattice has the clustering coefficient equal to zero), which is a property that has been found crucial for the evolution of cooperation, especially by games governed by pairwise interactions \cite{szabo_pr07}. In particular, while interaction graphs with zero or negligible clustering coefficients are characterized by an optimal level of uncertainty at which cooperators can survive, as was reported in \cite{vukov_pre06,perc_njp06a}, interaction graphs with overlapping triangles (which are inherent to the triangular lattice, as can be observed from figure~\ref{graphs}) preclude such an observation, \textit{i.e.} the $D \to C+D$ phase boundary is monotonically descending towards lower $r$ in the $K \to 0$ limit. Remarkably, recent results for games governed by group interactions indicate that there the interaction graph may play a less crucial role, since in fact joint memberships in the same groups effectively link players that are otherwise not directly connected \cite{szolnoki_pre09c}. By comparing the results presented in figures~\ref{sqr} and \ref{tri} for the same type of heterogeneity in $h_x$, this is indeed fully confirmed as all the main features are identical. Altogether, these results solidify the impression that strongly heterogeneous states in the public goods game can be less effective in warranting high levels of cooperation than moderately heterogeneous states.

\begin{figure}
\begin{center} \includegraphics[width = 12cm]{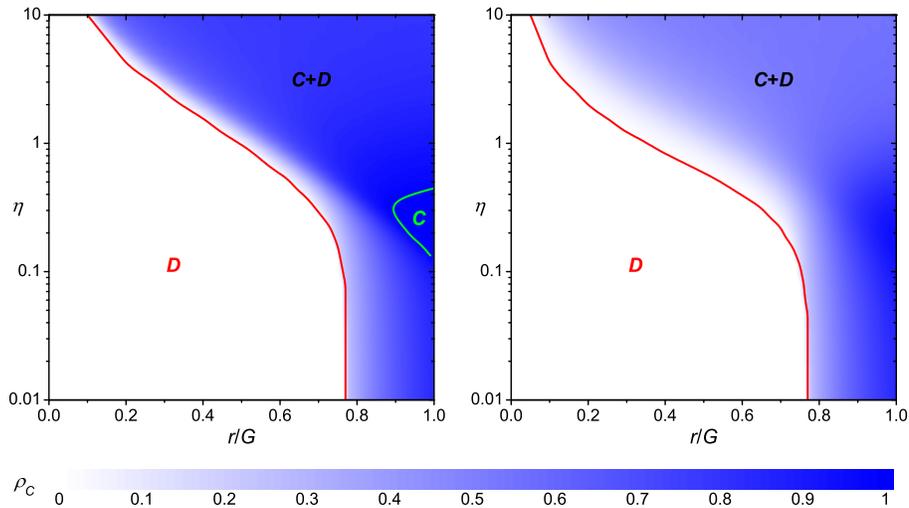}
\caption{\label{rrg} Full $r-\eta$ phase diagrams for the random regular graph with uniformly (left) and exponentially (right) distributed public goods. Red and green lines delineate the $D \to C+D$ and $C+D \to C$ phase transitions, respectively. The blue shading corresponds to the stationary density of cooperators $\rho_C$ at each particular combination of $r$ and $\eta$, as depicted by the color bar at the bottom. Lack of spatial structure, if compared to the square and the triangular lattice, also imposes the same conclusion, namely that uniformly distributed public goods (left) promote the evolution of cooperation better than exponentially distributed public goods (right).}
\end{center}
\end{figure}

As the last type of interaction graph we consider the random regular graph (RRG), which unlike the square lattice and the triangular lattice, lacks spatial structure. Note that locally the random regular graph is similar to the Bethe tree (see figure~\ref{graphs} right). Since the seminal paper by Nowak and May \cite{nowak_n92b} clearly established spatial structure as an important agonist for the successful evolution of cooperation, it is certainly appropriate to test our results also against this variation. Moreover, due to the lack of spatial structure, there is an expected difference in the directed percolation universality class \cite{odor_rmp04} if compared to the square lattice. Results presented in figure~\ref{rrg} depict full $r-\eta$ phase diagrams for uniformly (left) and exponentially (right) distributed public goods. Compared to results obtained for the square lattice (see figure~\ref{sqr}), on the RRG too the position of the $D \to C+D$ transition line remains largely unaffected by the type of distribution of $h_x$. Moreover, the mixed $C+D$ phase for exponentially distributed public goods (right panel of figure~\ref{rrg}) is characterized by substantially lower values of $\rho_C$ (brighter shades of blue prevail) and the pure $C$ region is altogether missing. It is thus from results obtained for the RRG that in fact the failure of strongly heterogeneous states to reach the same levels of cooperation promotion as moderately heterogeneous states is most obvious. At the same time, as already noted above for the square as well as the triangular lattice, the position of the $D \to C+D$ transition line does not shift due to the difference in the distribution of $h_x$, which highlights that in games governed by group interactions the details of the interaction graph play a less prominent role than this is the case for games governed by pairwise interactions. With ample support from these final observations, we conclude that the strongly heterogeneous distribution is a less potent promoter of cooperative behavior in the public goods game.

\section{Summary}

We have studied the evolution of cooperation on three different types of regular graphs in the public goods game with uniformly and exponentially distributed payoffs, with the aim of determining to what extent do strongly heterogeneous states promote cooperative behavior. The presented results have important implications for the evolution of cooperation in games governed by group interactions. While past research has undoubtedly shown that strongly heterogeneous states facilitate the evolution of cooperation in games governed by pairwise interactions (for reviews see \cite{szabo_pr07,perc_bs10}), results presented in this paper question this in relation to moderate heterogeneities for games governed by group interactions. In fact, here we find that uniformly distributed public goods are more successful in maintaining high levels of cooperation than exponentially distributed public goods, which is different to what was reported previously for games governed by pairwise interactions \cite{perc_pre08}. This conclusion prevails irrespective of the type of the interaction graph, and is in agrement with previous observations \cite{szolnoki_pre09c} in that qualitative differences between the outcomes of games governed by pairwise interactions and games governed by group interaction should be expected due to the indirect linkage of players that are members of the same groups. A direct consequence of this indirect linkage is also the fact that differences in the clustering coefficient play a minor role by the promotion of cooperation, which is different from what was reported previously for games governed by pairwise interactions \cite{assenza_pre08}, where a high clustering coefficient was found to be tightly linked to flourishing cooperative states. Accordingly, in our case one could expect the triangular lattice to promote cooperation much better than the square lattice or the random regular graph, yet the difference is quite marginal. Moreover, it is also important to note that the random regular graph is fundamentally different from the square and the triangular lattice in that it has no local structure. The indirect linkage of players due to group interactions, however, renders these differences in structural properties of the interaction networks virtually irrelevant. Altogether, results presented in this paper point clearly towards the fact that the transition from pairwise to group interactions has important implications for the successful evolution of cooperation. We hope that this work will inspire future studies aimed at clarifying the role of heterogeneity for other types of games and further promote quantitative research in social systems with methods from physics \cite{castellano_rmp09}.

\ack
We thank Eberhard Bodenschatz for valuable comments towards the presentation of our work, as well as Attila Szolnoki and Gy{\"o}rgy Szab{\'o} for illuminating discussions and helpful advice. This research was supported by the Slovenian Research Agency (grants Z1-2032 and J1-4055).

\section*{References}
\providecommand{\newblock}{}

\end{document}